\documentstyle[prl,preprint,aps,epsf]{revtex}
\begin{document}

\title{Analytical Bounce Solution in a Dissipative Quantum Tunneling} 
%\preprint{MCTP-01-19}
\author{
D. K. Park\footnote{Email: dkpark@hep.kyungnam.ac.kr 
} }
\address{Department of Physics, Kyungnam University, Masan, 631-701, Korea}

\maketitle

%\date{\today}
\maketitle
\begin{abstract}
The analytical bounce solution is derived in terms of the polygamma function
in the Caldeira-Leggett's dissipative quantum tunneling model. The classical
action for the bounce solution lies between the upper and lower bounds in
the full range of $\alpha$, where $\alpha$ is a dissipation coefficient. The
bounce peak point increases from $1$ to $4/3$ with increase of $\alpha$. In
spite of various nice features we have shown that the solution we have derived
is not exact one by making use of the zero mode argument in the linearized
fluctuation equation. However, our solution can be a starting point for  
approximate computation of the prefactor in this model.
\end{abstract}

%---------------------------------------------------------------------------
\newpage
%\section{Introduction}
How to describe a dissipation at the level of 
quantum mechanics is a long-standing puzzle in physics. 
Upon our knowledge Feynman and Vernon(FV)\cite{feyold} 
firstly
described it as an interaction between a system of interest and its 
enviroments. Especially, they developed a formalism to investigate the 
quantum dissipation systematically by introducing an influence functional. The 
influence functional is an extremely important quantity in the sense that it 
contains all quantum effects of the enviroments and, thus makes it possible 
to describe the quantum dissipation in terms of only system's  coordinates. 

Based on FV formalism Caldeira and Leggett(CL) considered in Ref.\cite{cald83} 
a quantum tunneling model interacting with the harmonic oscillator enviroments. 
Introducing a potential which does not have a true vacuum, CL examined the 
effect of dissipation on quantum tunneling within a semi-classical 
approximation which was  
developed several decades ago\cite{col77,call77,raja82}. The
full application of the semi-classical application, however, is very difficult
in this setup because the exact bounce solution is still unknown. In fact, it 
seems to be extremely hard (or may be impossible) to derive a bounce solution
in an analytic form due to the non-local term the CL model involves. 
Without the analytic bounce
it is impossible to exploit the power of the semi-classical approximation
maximally. 

In Ref.\cite{cald83} CL obtained the upper and lower bounds of the 
classical action in terms of the dissipation coefficient $\alpha$ without
the explicit bounce solution. 
These two bounds are monotonically increasing function
with respect to $\alpha$, which indicates that the presence of the dissipation
causes the decrease of the tunneling probability within an exponential
approximation. The similar physical setup with a double-well potential
was examined by a 
canonical method in Ref.\cite{fuji92-1,fuji92-2}. The authors in 
Ref.\cite{fuji92-1,fuji92-2}, however, claimed that the dissipation may 
enhance the tunneling
probability. In order to reconcile these two apparently 
discrepant  results we think
the prefactor should be examined in the semi-classical side. However, it is 
very difficult to compute the prefactor without the analytic bounce solution.
Thus we may need a bounce solution in the analytic form  
for the computation of the prefactor although it is not exact. 
It is main purpose of this
letter to derive an analytic bounce which interpolates between 
the exact no-damping
solution and the strong-damping solution. 

We start with a dimensionless action\cite{cald83}\footnote{In this letter we 
will follow the same conventions with Ref.\cite{cald83}}
\begin{equation}
\label{action1}
\sigma[z] = \int_{-\infty}^{\infty} du
\left[ \left( \frac{d z}{d u} \right)^2 + (z^2 - z^3) \right]
+ \frac{\alpha}{\pi}
\int_{-\infty}^{\infty} du \int_{-\infty}^{\infty} du'
\left(\frac{z(u) - z(u')}{u - u'} \right)^2
\end{equation}
which yields an equation of motion as 
\begin{equation}
\label{eqmotion}
\ddot{z} = z - \frac{3}{2} z^2 + \frac{2 \alpha}{\pi}
\int_{-\infty}^{\infty} du' \frac{z(u) - z(u')}{(u - u')^2}.
\end{equation}
We know the exact no-damping ($\alpha \rightarrow 0$) solution
\begin{equation}
\label{no-damping}
z_0(u) = \mbox{sech}^2 \frac{u}{2}
\end{equation}
and strong-damping ($\alpha \rightarrow \infty$) solution
\begin{equation}
\label{strong-damping}
z_{\infty}(u) = \frac{4}{3 [ 1 + \left(\frac{u}{2 \alpha}\right)^2 ]}.
\end{equation}
Thus, the real bounce solution should interpolates between (\ref{no-damping})
and (\ref{strong-damping}) with increasing the dissipation coefficient
$\alpha$. 

The bounce solution we obtained in this letter is 
\begin{equation}
\label{main1}
z(u) = \sqrt{\frac{2}{\pi}} \frac{A_{\alpha}}{4 B_{\alpha}^2}
\left[ \psi' \left( \frac{C_{\alpha} + B_{\alpha} + i u}{2 B_{\alpha}} \right)
+ \psi' \left( \frac{C_{\alpha} + B_{\alpha} - i u}{2 B_{\alpha}} \right)
                                                                  \right]
\end{equation}
where $\psi'$ is an usual polygamma function. The $A_{\alpha}$, $B_{\alpha}$, 
and $C_{\alpha}$ are $\alpha$-dependent but $u$-independent constants which
obey the following equations:
\begin{eqnarray}
\label{const1}
& &\mu \psi''(\mu) + 2 \psi'(\mu) = \frac{2 \sqrt{2 \pi}}{3}
                                \frac{B_{\alpha}^2}{A_{\alpha}}
                                                           \\   \nonumber
& &\mu^2 \psi''(\mu) + 2 \mu \psi'(\mu) = \frac{4 \sqrt{2 \pi} \alpha}{3}
                                       \frac{B_{\alpha}}{A_{\alpha}}
                                                           \\   \nonumber
& &(3 - 8 \alpha^2) \mu^3 \psi''(\mu) + 6 (1 - 4 \alpha^2) \mu^2 \psi'(\mu)
+ 16 \alpha^2 \mu + 4 \alpha^2 = 0
\end{eqnarray}
where $\psi''$ is polygamma function and $\mu = C_{\alpha} / B_{\alpha}$. From
the first and second equations of Eq.(\ref{const1}) one can show easily 
$C_{\alpha} = 2 \alpha$. One can show also numerically 
that $A_{\alpha}$ and $B_{\alpha}$
are monotonically increasing functions with respect to $\alpha$. 

Before we explain how Eq.(\ref{main1}) is derived, we would like to show
its nice features. Fig. 1 shows the $\alpha$-dependence of its classical 
solution which is represented by the red line. The green and blue lines 
represent the classical actions for the strong-damping and no-damping
solutions respectively. Two black lines are upper and lower bounds of the 
classical action which were derived explicitly in 
Ref.\cite{cald83}. Fig. 1 indicates that the classical action for our bounce
solution (\ref{main1}) lies between the upper and lower bounds in the full
range of $\alpha$. Fig. 1 also implies that our solution interpolates
between the no-damping and the strong-damping solutions. 
The small difference between the red and green lines at the large $\alpha$
region indicates that our solution is an approximate analytical solution. This
fact will be proven later by making use of the zero mode argument in the
fluctuation equation level.
Fig. 2 shows the $\alpha$-dependence of $z(0)$, which means
the peak point of the bounce. As CL predicted in Ref.\cite{cald83}, the 
peak point increases from $1$ to $4/3$ with increase of $\alpha$. 
 
Now, let me explain how the bounce solution (\ref{main1}) is derived. Taking a 
Fourier transform from $z(u)$ to $\tilde{z}(\omega)$, one can change the 
equation of motion (\ref{eqmotion}) in terms of $\tilde{z}(\omega)$ as 
following
\begin{equation}
\label{eqmotion2}
(\omega^2 + 2 \alpha |\omega| + 1) \tilde{z}(\omega) = 
\frac{3}{2 \sqrt{2 \pi}} \int_{-\infty}^{\infty} d \Omega
\tilde{z}(\omega - \Omega) \tilde{z}(\Omega).
\end{equation}
The explicit form of $\tilde{z}_0(\omega)$ and $\tilde{z}_{\infty}(\omega)$
which are Fourier transform of $z_0(u)$ and $z_{\infty}(u)$ become
\begin{eqnarray}
\label{dampings}
\tilde{z}_0(\omega)&=&2 \sqrt{2 \pi} \frac{\omega}{\sinh \pi \omega}
                                                            \\  \nonumber
\tilde{z}_{\infty}(u)&=& \frac{4 \sqrt{2 \pi} \alpha}{3} e^{-2\alpha |\omega|}.
\end{eqnarray}
As expected $\tilde{z}_0(\omega)$ and $\tilde{z}_{\infty}(\omega)$ are 
solutions of Eq.(\ref{eqmotion2}) without second term and first term in the 
left-handed side of Eq.(\ref{eqmotion2}) respectively. 
These are properties of no-damping
and strong-damping in $\omega$-space. It is worthwhile noting the 
$\omega$-dependence of $\tilde{z}_0(\omega)$ and $\tilde{z}_{\infty}(\omega)$.
Although $\omega$ is a dimensionless quantity, it should have a dimension
if we go back to the CL's original model which has a dimension. Thus, from the 
dimensional consideration in Eq.(\ref{eqmotion2}) we can conjecture that
$\tilde{z}(\omega)$ should have a same dimension with $\omega$ in the 
original theory except the strong-damping solution, which should be 
dimensionless. From this point of view we can understand the 
$\omega$-dependence of $\tilde{z}_0(\omega)$ and $\tilde{z}_{\infty}(\omega)$.
Thus one can take an {\it ansatz}
\begin{equation}
\label{ansatz}
\tilde{z}(\omega) = A_{\alpha} \frac{\omega}{\sinh B_{\alpha} \omega}
e^{-C_{\alpha} |\omega|}.
\end{equation}
Then the $\alpha$-dependent constants $A_{\alpha}$, $B_{\alpha}$, and 
$C_{\alpha}$ should satisfy $A_0 = 2 \sqrt{2 \pi}$, $B_0 = \pi$, 
$C_0 = 0$, $A_{\infty} / B_{\infty} = 4 \sqrt{2 \pi} \alpha / 3$ and 
$C_{\infty} = 2 \alpha$ in order for $\tilde{z}(\omega)$ to interpolate 
between the 
no-damping and the strong-damping solutions.

Now, we will insert the {\it ansatz} (\ref{ansatz}) into Eq.(\ref{eqmotion2}) 
to extract an information on $A_{\alpha}$, $B_{\alpha}$, and $C_{\alpha}$.
The most difficult term we need to compute is the following convolution
term
\begin{equation}
\label{convolution}
\tilde{I}_{\alpha} (\omega) = \int_{-\infty}^{\infty} d \Omega
\tilde{z}(\omega - \Omega) \tilde{z}(\Omega)
= \frac{A_{\alpha}^2}{2}
\left[\tilde{I}_{1,\alpha}(\omega) - \omega^2 \tilde{I}_{2,\alpha}(\omega)
                                                               \right]
\end{equation}
where
\begin{eqnarray}
\label{defI12}
\tilde{I}_{1,\alpha}(\omega)&=& \int_0^{\infty} dy 
\frac{y^2}{\cosh B_{\alpha} y - \cosh B_{\alpha} \omega}
e^{-\frac{C_{\alpha}}{2} (|\omega + y| + |\omega - y|)}
                                                      \\  \nonumber
\tilde{I}_{2,\alpha}(\omega)&=& \int_0^{\infty} dy
\frac{1}{\cosh B_{\alpha} y - \cosh B_{\alpha} \omega}
e^{-\frac{C_{\alpha}}{2} (|\omega + y| + |\omega - y|)}.
\end{eqnarray}
Note that $\tilde{I}_{1,\alpha}(\omega)$ and $\tilde{I}_{2,\alpha}(\omega)$ are
even function with respect to $\omega$. Thus we can assume $\omega > 0$
without loss of generality. With this assumption 
$\tilde{I}_{1,\alpha}(\omega)$ and $\tilde{I}_{2,\alpha}(\omega)$ are expressed
as following
\begin{eqnarray}
\label{exprI12}
\tilde{I}_{1,\alpha}(\omega)&=& \frac{2}{3} 
\left[ \left(\frac{\pi}{B_{\alpha}}\right)^2 - \frac{\omega^2}{2} \right]
\frac{\omega}{\sinh B_{\alpha} \omega} + 
\frac{1}{B_{\alpha}^3}
\left( \partial_{\mu}^2 K - e^{-\mu z_0}
       \partial_{\mu}^2 K \bigg|_{\mu = 0} \right)
                                                     \\  \nonumber
\tilde{I}_{2,\alpha}(\omega)&=& -\frac{\omega}{\sinh B_{\alpha} \omega}
e^{-C_{\alpha} \omega} + \frac{1}{B_{\alpha}}
\left( K - e^{-\mu z_0} K\bigg|_{\mu = 0} \right)
\end{eqnarray}
where
\begin{equation}
\label{defKK}
K \equiv \int_{z_0}^{\infty} dz 
\frac{e^{-\mu z}}{\cosh z - \cosh z_0}
\end{equation}
and $z_0 \equiv B_{\alpha} \omega$ and $\mu \equiv C_{\alpha} / B_{\alpha}$.
When we compute the first term of $\tilde{I}_{1,\alpha}(\omega)$ we used the
property of the Lerch function in Ref.\cite{grad00}.

Now, the remaining problem for the computation of $\tilde{I}_{\alpha} (\omega)$
is to compute $K$ which have an infrared-like infinity 
as a field theory
terminology. In order to take into account the infinity carefully we take 
a change of variable $x=e^z$, which makes $K$ to be
\begin{equation}
\label{reexprK}
K = 2 (x_0 - x_0^{-1})^{-1}
\int_{x_0 + \epsilon}^{\infty} dx
\left[ \frac{x^{-\mu}}{x - x_0} - \frac{x^{-\mu}}{x - x_0^{-1}} \right]
\end{equation}
where $x_0 = e^{z_0}$. In Eq.(\ref{reexprK}) we introduced an infinitesimal 
parameter $\epsilon$ explicitly for the regularization of the infrared-like
infinity. Performing the integration in Eq.(\ref{reexprK}) one can
express $K$ as a difference of two hypergeometric functions. Making use
of the relation between the hypergeometric and digamma function\cite{abra72}
the final expression of $K$ becomes
\begin{equation}
\label{finalK}
K = \frac{e^{-\mu z_0}}{\sinh z_0}
\left[(z_0 - \ln \epsilon) + e^{2 \mu z_0} \ln (e^{2 z_0} - 1)
      + (e^{2 \mu z_0} - 1) \psi (\mu)
      - \sum_{n=1}^{\infty} \frac{(\mu)_n}{n!} \psi(n+1) (1 - e^{-2 z_0})^n
                                                           \right]
\end{equation}
where $(\mu)_n = \mu (\mu + 1) \cdots (\mu + n - 1)$ and $\psi$ is a digamma
function. Note that $K$ has a logarithmic divergence as expected. Using 
Eq.(\ref{finalK}) it is straightforward to compute 
$\tilde{I}_{2,\alpha}(\omega)$ which reduces to
\begin{eqnarray}
\label{i2alpha}
\tilde{I}_{2,\alpha}(\omega)&=& \frac{e^{-C_{\alpha} \omega}}{B_{\alpha}
                               \sinh B_{\alpha} \omega}
\Bigg[ z_0 + \left(e^{2 \mu z_0} - 1\right)
      \left\{\psi(\mu) + \ln \left(e^{2 z_0} - 1\right) \right\}
                                                              \\  \nonumber
& & \hspace{5.0cm}
-\sum_{n=1}^{\infty} \frac{(\mu)_n}{n!} \psi(n+1)
 \left(1 - e^{-2 z_0}\right)^n \Bigg].
\end{eqnarray}
Note that the infinity term in Eq.(\ref{finalK}) disappears in 
Eq.(\ref{i2alpha}) because of the exact cancellation. This exact cancellation 
also takes place in $\tilde{I}_{1,\alpha}(\omega)$. After tedious calculation
the final form of $\tilde{I}_{\alpha}(\omega)$ reduces to
\begin{eqnarray}
\label{longg}
& &\tilde{I}_{\alpha}(\omega) \tilde{z}^{-1}(\omega) = \frac{1}{3} A_{\alpha}
\left[\omega^2 + \left(\frac{\pi}{B_{\alpha}}\right)^2\right]
                                                              \\  \nonumber
& & \hspace{1.5cm}
+ \frac{A_{\alpha}}{2 B_{\alpha}^2}
\Bigg[\left\{\frac{e^{2 \mu z_0} - 1}{z_0} \psi''(\mu) + 
      2 \left(e^{2 \mu z_0} + 1\right) \psi'(\mu) \right\} - 
      \left(\frac{4}{3} z_0^2 + \frac{2 \pi^2}{3} \right)
                                                              \\  \nonumber
& & \hspace{3.0cm}
      + 2 \sum_{n=1}^{\infty} \frac{(\mu)_n}{n!} 
        \left[ \psi(n + \mu) - \psi(\mu) \right] \psi(n+1) 
       \left(1 - e^{-2 z_0}\right)^n                         
                                                              \\   \nonumber
& & \hspace{3,0cm} 
 -2 \sum_{n=1}^{\infty} \frac{\psi(n+1)}{n} \left(1 - e^{-2 z_0}\right)^n
                                                               \\   \nonumber
& &   \hspace{3.0cm}
  - \sum_{n=2}^{\infty} \frac{(\mu)_n}{n!}
   \left( [\psi'(n+\mu) - \psi'(\mu)] + [\psi(n+\mu) - \psi(\mu)]^2 \right)
    \psi(n+1) \frac{\left(1 - e^{-2 z_0}\right)^n}{z_0}
                                                               \\  \nonumber
& & \hspace{3.5cm}
    + 2 \sum_{n=2}^{\infty} \frac{\gamma + \psi(n)}{n} \psi(n+1)
      \frac{\left(1 - e^{-2 z_0}\right)^n}{z_0}   
                                                    \Bigg]
\end{eqnarray}
where $\gamma$ is an Euler's constant. In order for $\tilde{z}(\omega)$ to be
an exact solution the right-handed side of Eq.(\ref{longg}) should be equal to
$2\sqrt{2 \pi} (\omega^2 + 2 \alpha \omega + 1) / 3$. To extract an information
on $A_{\alpha}$, $B_{\alpha}$, and $C_{\alpha}$ 
we assume this equality. Repeating
to take a 
$\omega \rightarrow 0$ limit and subsequently to differentiate the 
right-handed side of Eq.(\ref{longg})
three times, we can derive Eq.(\ref{const1}). Taking an inverse-Fourier 
transform to $\tilde{z}(\omega)$, we can derive Eq.(\ref{main1}).

Although our bounce solution has many nice features as discussed before, it 
is not an exact solution unfortunately except $\alpha = 0$. For $\alpha = 0$
case  we can 
show analytically $A_0 = 2 \sqrt{2 \pi}$, $B_0 = \pi$ and $C_0 = 0$ by solving
Eq.(\ref{const1}) in the $\mu \rightarrow 0$ limit. This means our solution 
exactly coincides with no-damping solution at $\alpha \rightarrow 0$ limit. 
However, for the nonzero $\alpha$ we can show that our bounce solution is an
approximate one by using the zero mode of the linearized fluctuation equation
as following. 

Note that the CL model has a time-translational symmetry in 
spite of the presence of the non-local term. Although one can show this fact
simply from the equation of motion (\ref{eqmotion}), we would like to show it 
at the fluctuation level for a later use. Inserting 
$z(u) = z_{cl}(u) + \eta(u)$ into Eq.(\ref{eqmotion}) one can construct easily
the linearized fluctuation equation
\begin{equation}
\label{fluc1}
-c_1 \ddot{\eta} + c_2 (1 - 3 z_{cl}) \eta + c_3 \frac{2 \alpha}{\pi}
\int_{-\infty}^{\infty} du'
\frac{\eta(u) - \eta(u')}{(u - u')^2} = \lambda \eta
\end{equation}
where $\lambda$ is an eigenvalue of the fluctuation equation and the 
constants $c_1$, $c_2$, and $c_3$ are introduced for convenience. If 
$c_1 = c_2 = 1$ and $c_3 = 0$, Eq.(\ref{fluc1}) is a fluctuation equation 
around the no-damping solution. In this case it is trivial to show that 
$dz_0(u) / du$ is a zero mode. If $c_1 = 0$ and $c_2 = c_3 = 1$, 
Eq.(\ref{fluc1}) corresponds to a fluctuation around the strong-damping 
solution. In this case also one can show directly that 
$dz_{\infty}(u) / du$ is a zero mode. This means that the CL model has a 
time-translational symmetry in spite of the presence of the dissipation. 

Now, let us consider the full fluctuation equation, {\it i.e.}
$c_1 = c_2 = c_3 = 1$. If our bounce solution (\ref{main1}) is an 
exact one, by same reason 
\begin{equation}
\label{fluc2}
\eta \equiv \frac{d z(u)}{d u} = \sqrt{\frac{2}{\pi}}
\frac{i A_{\alpha}}{8 B_{\alpha}^3}
\left[ \psi'' \left(\frac{C_{\alpha} + B_{\alpha} + i u}{2 B_{\alpha}}\right)
      - \psi'' \left(\frac{C_{\alpha} + B_{\alpha} - i u}{2 B_{\alpha}}\right) 
                                                                \right]
\end{equation}
should be a zero mode. Inserting Eq.(\ref{fluc2}) into the non-local term in
Eq.(\ref{fluc1}) one can show 
\begin{equation}
\label{nonlocal1}
\frac{2 \alpha}{\pi} \int_{-\infty}^{\infty} du'
\frac{\eta(u) - \eta(u')}{(u - u')^2} = 
\sqrt{\frac{2}{\pi}} \frac{3 i \alpha A_{\alpha}}{4 B_{\alpha}^4}
\left[\zeta\left(4, \frac{C_{\alpha} + B_{\alpha} - i u}{2 B_{\alpha}}\right) -
      \zeta\left(4, \frac{C_{\alpha} + B_{\alpha} + i u}{2 B_{\alpha}}\right)
                                                              \right]
\end{equation}
where $\zeta(p, q)$ is a Riemann Zeta function defined as 
$\zeta(p,q) = \sum_{k=0}^{\infty} 1 / (q+k)^p$.

Using Eq.(\ref{nonlocal1}) the left-handed side of the fluctuation can be 
plotted 
numerically. When $\alpha$ is small, the left-handed side 
of the fluctuation equation is 
plotted in Fig. 3 in terms of $u$ for various $\alpha$, 
which indicates that $\eta$ is not exact
zero mode although approximately it is. Fig. 3 also shows how $\eta$ can be a 
zero mode in the $\alpha \rightarrow 0$ limit. Fig. 4 is a plot of the 
left-handed side of 
Eq.(\ref{fluc1}) when $\alpha$ is large. This figure also shows how our 
solution goes to the strong-damping solution in the $\alpha \rightarrow \infty$
limit. But we should comment that our bounce solution does not seem to 
exactly coincide
with the strong-damping solution in the $\alpha \rightarrow \infty$
limit. That is why there is a small difference in Fig.1 
between the classical action for our solution and its lower bound
at the large $\alpha$ regime. 

In this letter we have derived the analytic bounce solution in the CL model.
The classical action for our solution lies between the upper and lower bounds
in the full range of $\alpha$. Although it has many nice features, we have
shown that it is not an exact solution except $\alpha = 0$ case. However, 
using this approximate solution, one may be able to compute the prefactor 
approximately. We guess this prefactor may be important factor to reconcile
the discrepancy between the semi-classical method and the canonocal method.
We hope to visit this issue in the near future. Our analytic bounce solution 
might be extended to the non-Ohmic dissipation case. The explicit result will
be discussed elsewhere in detail.

\vspace{1cm}

{\bf Acknowledgement}:  
This work was support by the Korea Research 
Foundation under Grant (KRF-2003-015-C00109).

%\newpage
%\begin{appendix}{\centerline{\bf Appendix A}}
%\setcounter{equation}{0}
%\renewcommand{\theequation}{A.\arabic{equation}}

%\end{appendix}

\begin{figure}
\caption{Plot of $\alpha$-dependence of classical actions for our bounce
solution (\ref{main1}) (red line), no-damping solution (\ref{no-damping})
(blue line) and strong-damping solution (\ref{strong-damping}) (green line).
Two black lines are upper and lower bounds of the classical action which were
derived by CL in Ref.[2] without an explicit bounce solution. 
This figure 
indicates that the classical action for our solution lies between the upper
and lower limits in the full range of $\alpha$.}  
\end{figure}
\vspace{0.4cm}
\begin{figure}
\caption{Plot of $\alpha$-dependence of the bounce peak point. As CL have 
argued in Ref.[2], the bounce peak point increases from $1$ to 
$4/3$ with increase of $\alpha$.} 
\end{figure}
\vspace{0.4cm}
\begin{figure}
\caption{The plot of the left-handed side of Eq.(\ref{fluc1}) for various 
small $\alpha$. The difference from zero indicates that our solution is not
an exact one. The increase of the peak point and the wide-spreading shape of 
peak with increase of $\alpha$ denotes that our bounce solution (\ref{main1}) 
goes away from 
the exact one when $\alpha$ becomes larger in the small
$\alpha$ regime.} 
\end{figure}
\vspace{0.4cm}
\begin{figure}
\caption{Plot of the left-handed side of Eq.(\ref{fluc1}) for various large
$\alpha$. The decrease of the peak point with increase of $\alpha$ means that 
our  solution approaches to the exact one when $\alpha$ becomes larger in the 
large $\alpha$ regime.}
\end{figure}

\newpage
\epsfysize=10cm \epsfbox{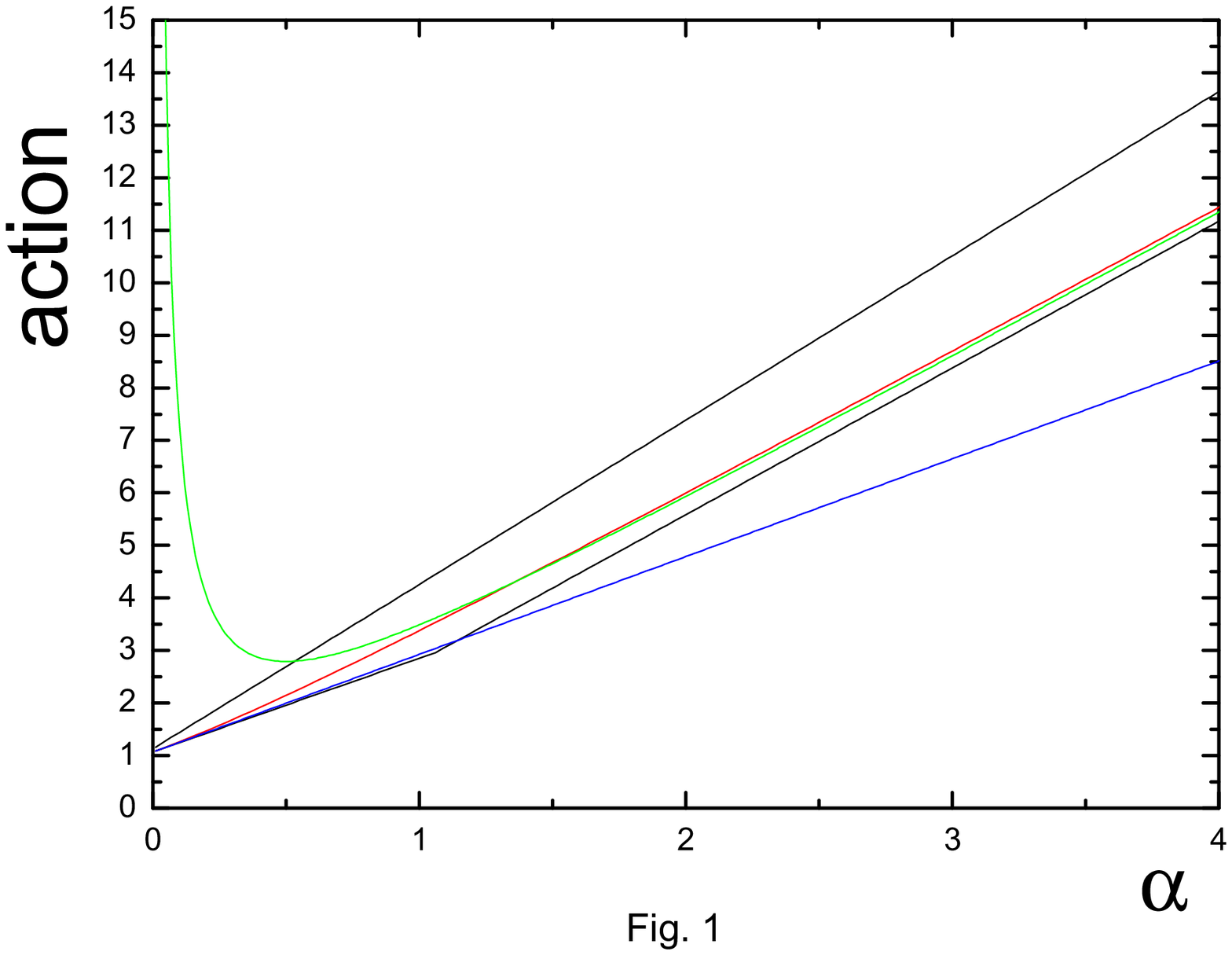}
\newpage
\epsfysize=10cm \epsfbox{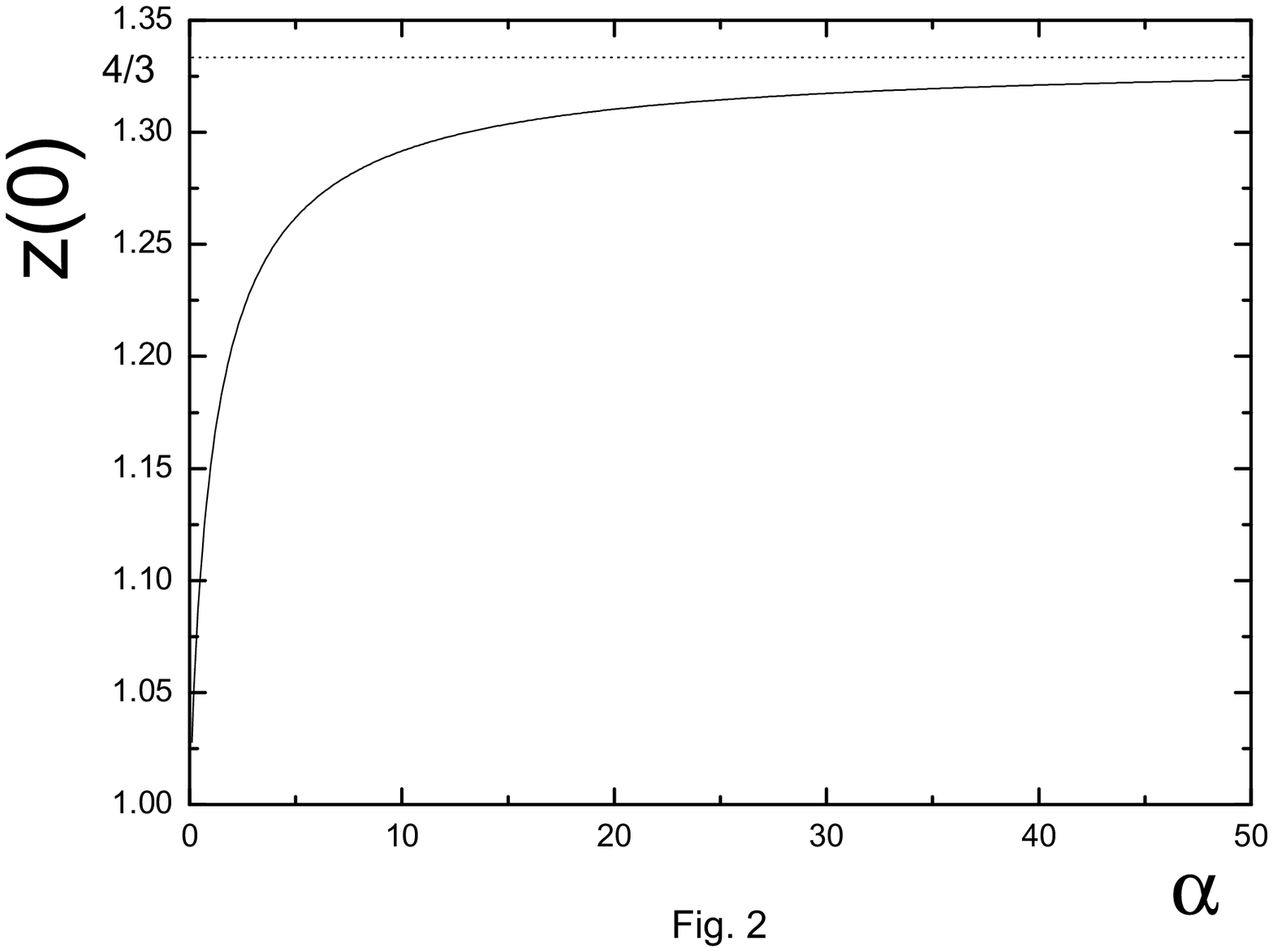}
\newpage
\epsfysize=10cm \epsfbox{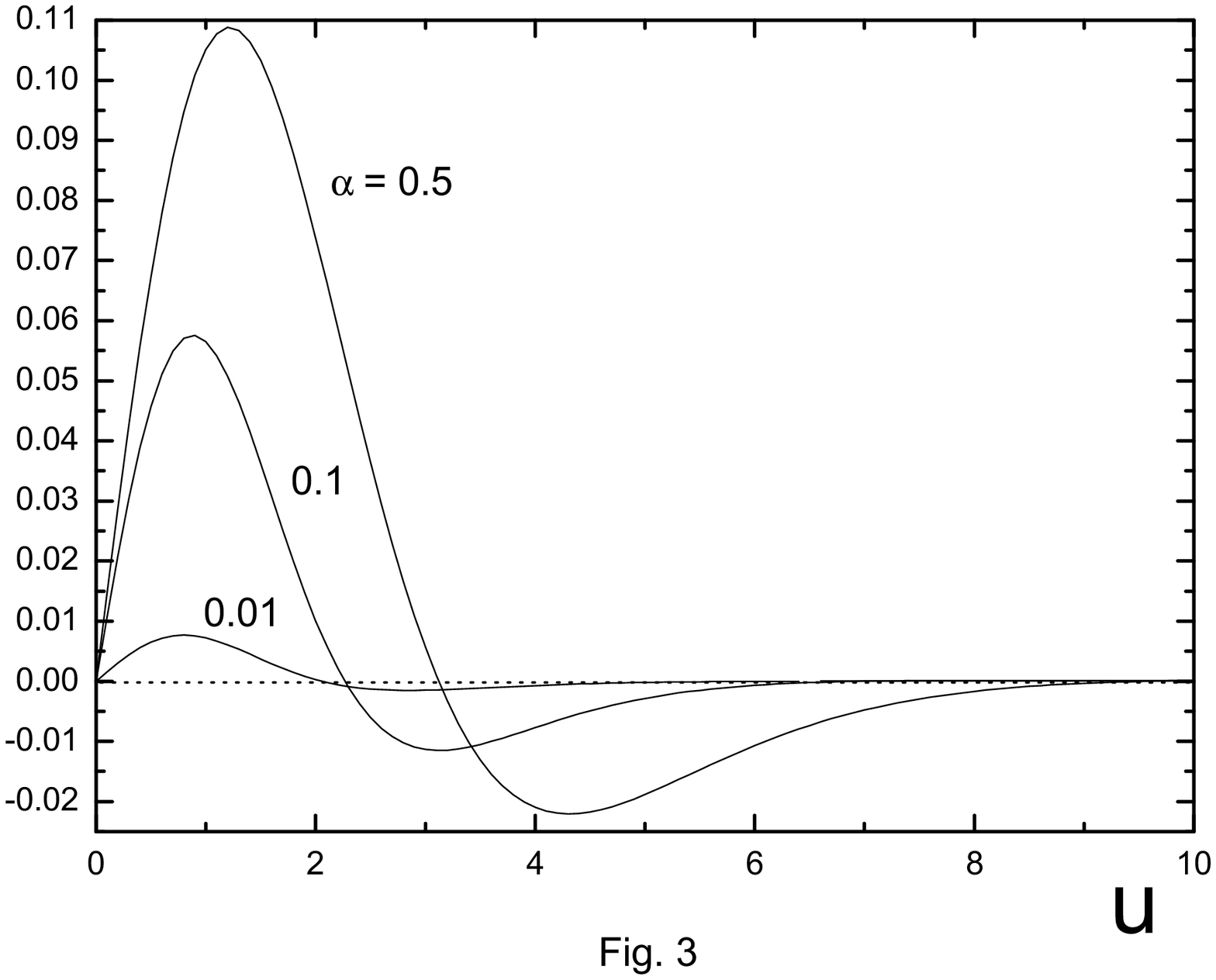}
\newpage
\epsfysize=10cm \epsfbox{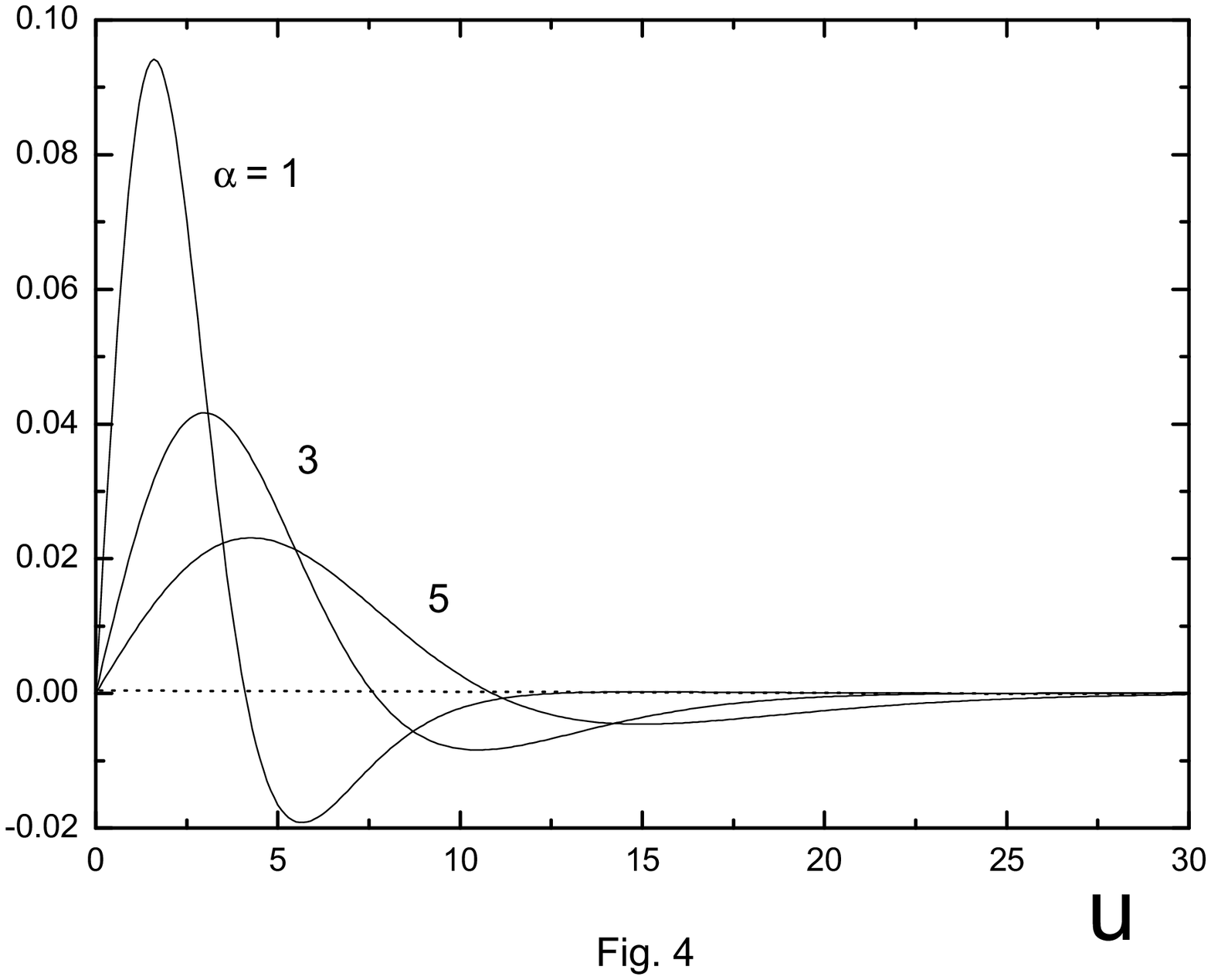}
%\newpage
%\epsfysize=20cm \epsfbox{fig5.ps}
%\newpage
%\epsfysize=20cm \epsfbox{fig6.ps}
\end{document}